\documentclass[pre,aps,twocolumn,superscriptaddress,pacs]{revtex4}
\usepackage{epsfig}
\usepackage{bm}
\newcommand{\B}[1]{{\bm{#1}}}

\usepackage[latin1]{inputenc}
\newcommand{\beq}{\begin{equation}}
\newcommand{\eeq}{\end{equation}}
\newcommand{\bea}{\begin{eqnarray}}
\newcommand{\eea}{\end{eqnarray}}
\begin{document}
\title{The Universality Class of Diffusion Limited Aggregation and Viscous Fingering}
\author{Joachim Mathiesen}
\affiliation{NTNU, Institutt for Fysikk, 7491 Trondheim, Norway}
\author{Itamar Procaccia}
\affiliation{The Department of Chemical Physics, The Weizmann
Institute of Science, Rehovot 76100, Israel}
\author{Harry L. Swinney}
\affiliation{Center for Nonlinear Dynamics and Department of
Physics, University of Texas at Austin, Texas 78712}
\author{Matthew Thrasher}
\affiliation{Center for Nonlinear Dynamics and Department of
Physics, University of Texas at Austin, Texas 78712}

\begin{abstract}
 We investigate whether fractal viscous fingering and diffusion limited
aggregates are in the same scaling universality class.  We bring
together the largest available viscous fingering patterns and a
novel technique for obtaining the conformal map from the unit circle
to an arbitrary singly connected domain in two dimensions.  These
two Laplacian fractals appear different to the eye; in addition, viscous
fingering is grown in parallel and the aggregates by a serial
algorithm. Nevertheless, the data strongly indicate that these two
fractal growth patterns are in the same universality class.

\end{abstract}
\maketitle

Laplacian fractals are paradigmatic examples for
the spontaneous growth of fractal patterns in natural systems. In
particular, two such examples have attracted an enormous amount of
interest: viscous fingering and diffusion limited aggregation (DLA).

\begin{figure}
\centering
\epsfig{width=.4\textwidth,file=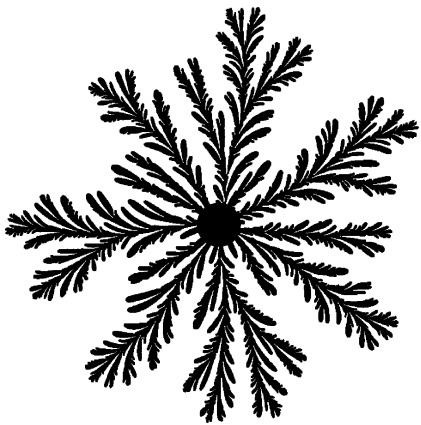,clip}
\caption{Digitized image of an experimental viscous fingering
pattern.  Air (black) is injected into the oil-filled gap (white).
 The pattern is approximately 22 cm in diameter, cf. \cite{05PS}.}
\label{visfig}
\end{figure}

Viscous fingering is realized \cite{86DNS} when a less viscous fluid
displaces a more viscous fluid contained in the narrow gap between two
glass plates (a Hele-Shaw cell). When the less viscous fluid is
inserted through an aperture in one of the glass plates, a pattern is
formed, as shown in Fig.~\ref{visfig} \cite{05PS}. The fluid velocity
$\B v$ of the displaced fluid in a Hele-Shaw cell satisfies Darcy's
law, $\B v\propto \B \nabla p$, where $p$ is the pressure. To a very
good approximation, the viscous fluid is incompressible (i.e., $\B
\nabla\cdot \B v =0$), so the pressure in the viscous fluid satisfies
the Laplace equation $\nabla^2 p=0$.

\begin{figure}
\centering
\epsfig{width=.35\textwidth,file=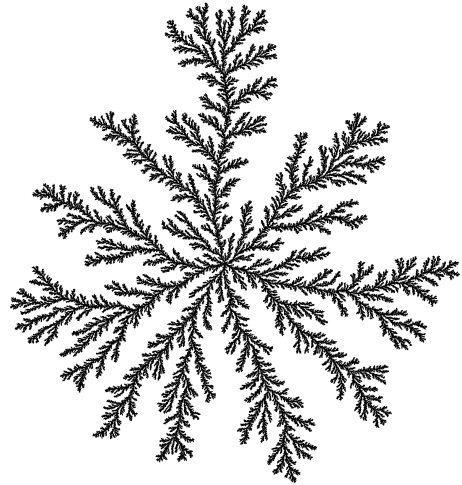}
\caption{A diffusion limited aggregate with 50,000,000 particles. Image courtesy of Ellak Somfai.} \label{DLA}
\end{figure}

DLA is realized \cite{81WS} as a computer experiment in which a
fractal cluster is grown by releasing a fixed size random walker from
infinity, allowing it to walk until it hits any particle already
belonging to the cluster. Since the particles are released one by one
and may take a long time to hit the cluster, the probability field is
stationary and, in the complement of the cluster, we have again
$\nabla^2 p=0$. A typical DLA cluster is shown in Fig. \ref{DLA}.

Mathematically, these problems are similar but not identical. In
both cases, one solves the Laplace equation with the same boundary
condition at infinity, i.e. $\B\nabla p =const\times \hat r/r$ as
$r\to \infty$. However, in viscous fingering, each point on the fractal's
boundary advances at a rate proportional to $\B\nabla p$, whereas
the DLA accretes one particle at a time, changing the Laplacian
field after each such growth step. Thus, we refer to viscous
fingering and DLA as parallel and serial processes, respectively. In
addition, the ultraviolet regularization differs; in DLA,  $p=0$ on
the cluster and the regularization is provided by the particle size.
In viscous fingering,  one solves the problem with the boundary
condition $p=\sigma\kappa$ where $\sigma$ is  the surface tension
and $\kappa$ is  the local curvature. Finally, viscous fingers are
grown in a finite gap and are not truly 2-dimensional. Accordingly,
one can ask whether  these two fractal growth problem are in the
same scaling universality class. But to answer this question, one
must first define what one means by a ``scaling universality
class.''

{\bf Definition of the scaling universality class}: The first test of
correspondence between the fractal patterns has to do with their
fractal dimension $D_0$.  Denote by $R_n$ the radius of the minimal
circle that contains a fractal pattern. The fractal dimension is
defined by how the mass $M_n$ contained within this circle (number of
particles for DLA, area for viscous fingers) scales with $R_n$,
$M_n\sim R_n^{D_0}$.  Measurements of this type indicated a value
$D_0\approx 1.71$ for both problems, motivating many authors to
express the opinion that these two problems {\em are} in the same
universality class \cite{05PS,see}. Obviously, the fractal dimension
by itself is not sufficient, and a more stringent definition is
necessary.

We propose here that the identity of the scaling properties of the
harmonic measure is a sufficient test for two Laplacian growth
problems to be in the same universality class. The harmonic measure
is the probability for a random walker to hit the boundary of the
fractal pattern. It determines the growth in both problems, being
proportional to $\B\nabla p$ at the boundary. Suppose that we know
the probability measure $\mu(s)ds$ for a random walker to hit an
infinitesimal arclength on the fractal boundary. We compute the
probability $P_i(\epsilon)\equiv \int_{i\textrm{'th box}} \mu(s)
ds$, and then define the generalized dimensions \cite{83HP} via
\begin{equation}
D_q \equiv \lim_{\epsilon\to 0} \frac{\log \sum_{i=1}^{N(\epsilon)}  P^q_i(\epsilon)}{(q-1)\log \epsilon} \ ,
\end{equation}
where $N(\epsilon)$ is the number of boxes of size $\epsilon$ that
cover the fractal boundary.  The $q>0$ branch of this
function probes the high probability region of the measure, whereas
the $q<0$ branch stresses the low probabilities. For $q\to 0^+$ we
find the fractal dimension $D_0$. An equivalently useful description
\cite{86HJKPS} is provided by the scaling indices $\alpha$ that
determine how the measure becomes singular, $P_i(\epsilon) \sim
\epsilon ^\alpha$, as $\epsilon \to 0$.  This set of indices is
accompanied by $f(\alpha)$ which is the fractal dimension of the
subset of singularities of scaling exponent $\alpha$. The relation
of these objects to $D_q$ is in the form of the Legendre transform
\begin{eqnarray}
\alpha(q)=\frac{\partial[(q-1)\, D_q]}{\partial q}\ , \\
f(\alpha) = q \,\alpha(q) - (q-1)\, D_q
\end{eqnarray}
In fractal measures defined on simple fractals, the values of $\alpha$
are usually bounded from above and from below, $\alpha_{min}\le
\alpha\le \alpha_{max}$. When the fractal measure fails to exhibit
power law scaling everywhere (say, $P_i(\epsilon)\sim e^{-\epsilon}$
somewhere), one finds a phase transition in this formalism \cite{87KP}
with one of the edge values $\alpha_{min}$ or $\alpha_{max}$ ceasing
to exist. {\em When the spectrum of exponents $\alpha$ and their
frequency of occurrence $f(\alpha)$ in two problems are the same, we
refer to these problems as being in the same universality class}. We
note that a similar criterion was used to state that different
dynamical systems are in the same universality class at their
transition to chaos \cite{85JLKPS}, but not many other physical
problems yielded to such a stringent test, simply because it is not
easy to compute with enough precision the scaling properties of
fractal measures. For DLA, such an accurate computation was achieved
\cite{01DJLMP}; in this Letter, we report also an accurate computation
for {\em experimental} viscous fingering patterns, allowing us to test
whether the two problems are in the same universality class.

The apparent complexity of the viscous fingers is discouraging for
any attempt to calculate the harmonic measure reliably.  The naive
method for computing the harmonic measure for DLA is to ``probe" the
interface with random walkers and perform frequency statistics
\cite{86ACL}. While the statistics at the outer tips is reasonable
\cite{86HMP},  the deep fjords are visited extremely rarely by the
random walkers. Similarly,  in viscous fingering experiments, the
measure is usually estimated from the velocity of the interface
\cite{87MBFJ}. Obviously, this reveals the outer tips which move
with an appreciable velocity, but leaves unmeasured the harmonic
measure on the fjords, which almost do not move at all. One has to
look for alternative methods.

{\bf The harmonic measure from iterated conformal maps}: The Riemann
theorem guarantees that there exists a conformal map from the exterior
of the unit circle $\omega=e^{i\theta}$ to the exterior of our viscous
fingering patterns. Having such a conformal map, say $\Phi(\omega)$,
the harmonic measure is simply obtained as $1/|\Phi'(\omega)|$ up to
normalization. The actual construction of such a map for a given
well-developed fractal pattern is, however, far from obvious, and it
has never been accomplished before.  We will demonstrate now that the
method of iterated conformal maps \cite{98HL,02JLMP} can be adapted to
solve this problem in a very efficient way.
\begin{figure}
\epsfig{width=.4\textwidth,file=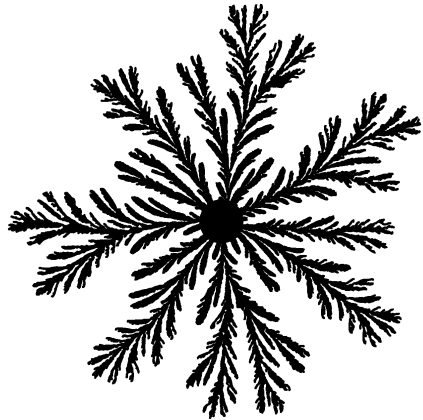}
\vspace{-.5cm}
\caption{ Conformal map of the unit circle aimed to fit the pattern in Fig.\ \ref{visfig}.
The map was constructed using the algorithm
  proposed in the text, transforming the unit disc represented by the black area. This figure demonstrates
  the accuracy of our method for reproducing complex patterns.}
  \label{map}
\end{figure}
We use data acquired in experiments, in which air is injected into
oil from a central orifice. The oil is confined between two circular
glass plates, 288 mm in diameter and each 60 mm thick \cite{05PS}.
The plates are separated by 0.127 mm and are flat to 0.13 $\mu$m
(optically polished to one-quarter of a wavelength). An oil buffer
surrounds the plates. The data presented here were obtained with
silicone oil with a dynamic viscosity $\mu$ = 345 mPa$\cdot$s and
surface tension $\sigma$=20.9 mN/m at 24$^\circ$C; additional
experiments were conducted with silicone oil with a dynamic
viscosity $\mu$ = 50.8 mPa$\cdot$s and surface tension $\sigma$=20.6
mN/m at 24$^\circ$C. The pressure difference $\Delta p$ between the
injected air and the oil buffer ranged from 0.1 to 1.25 atm. The 1
mm diameter central hole through which the air is injected cannot be
seen in Fig.~\ref{visfig}, rather the 22 mm diameter central solid
circular region (black) is a mask introduced after the images have
been digitized. The interface near the injection point could not be
extracted because of light contrast limitations.

The images are digitized and each pixel is given an index $j$ from
$1$ to $N$.   As a first step,  we rescale all the pixels and apply
a uniform shift such that the central black region becomes the unit
circle. Next, we record the position of each pixel in the complex
plane, denoting it by $z_j\equiv x_j+iy_j$.  Next, we sort the
pixels in ascending order with respect to $|z_j|$, which is the
minimal distance $r_j$ of the $j$'th pixel from the interface of the
fractal pattern.  We divide the sorted pixels into $m$ groups, such
that the $k$'th group contains the pixels which have absolute values
smaller than those in the $(k+1)$'th group. Within the individual
groups, we once again divide the pixels into two subgroups, where
the first subgroup contains the pixels with associated radii $r_j$
larger than a threshold value $\delta$. Typically, we choose
$\delta$ to be five times the pixel size. We build the conformal
mapping from the unit circle to the viscous fingers by expanding
stepwise the unit circle in the interior part of the viscous
fingers. The size of the expansion is given by $r_j$. In other
words, in the $n$'th step we build a conformal mapping
$\Phi^{(n)}(w)$ from the $\omega$ complex plane to the  $z$ complex
plane. At each step, we employ an auxiliary mapping
$\varphi_{\theta,\lambda}$ that locally perturbs the unit circle
with a semicircular bump centered at the position $e^{i\theta}$ and
of linear size $\sqrt \lambda$. The construction is done such that
we first expand around the points in the first subgroup of the first
group, then we proceed to the second subgroup. Once we have finished
expanding around the first group, we move on to the second group,
etc.
\begin{figure}
\centering
\epsfig{width=.4\textwidth,file=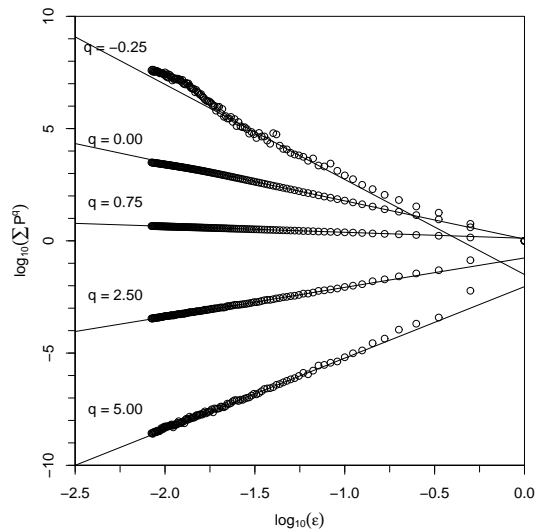}
\caption{Log-log plots of $\sum_{i=1}^{N(\epsilon)}  P^q_i(\epsilon)$ vs.
the box size $\epsilon$ for selected values of $q$. We find reliable power law scaling
for $q\ge -0.25$.}
\label{scaling}
\end{figure}
Most pixels in the vicinity of the pixel $z_j$ will be covered when
expanding the conformal mapping around $z_j$; therefore, they can be
disregarded in all future iterations.  In fact, a majority of the
pixels will be covered by expansions around nearby points. Thus we
need to perform a significantly smaller number of expansions than
the number of pixels. The aforementioned division of the pixels is a
simple way to speed up our method, i.e.\ to reduce the number of
necessary expansions. We first expand the mapping around the points
$\chi^{(n)}(z_j)$ with large radii ($r_j>\delta$) and close to the
unit circle. The conformal mapping is constructed by successive
iterations. In each iteration step, the mapping covers more and more
of the viscous fingers, and the number of pixels not being covered
decreases.

Assume that we have reached the $n$'th iteration step. Considering
the pixels that are not covered, we now use the inverse mapping,
$\chi^{(n)}=\left (\Phi^{(n)}\right ) ^{-1}$, to map these pixels
$z_j$ to the exterior of the unit circle, i.e.\ we map them from the
$z$-plane to the $w$-plane. The size $\sqrt{\lambda_{n+1}}$ of the
($n+1$)'th bump to be centered at the position $z_j$ is taken to be
\begin{equation}
\sqrt{\lambda_{n+1}}= \frac {r_j}{\left |\Phi^{'(n)}(e^{i\theta_{n+1}})\right |},
\mbox{ where}\quad \theta_{n+1}=\arg \chi^{(n)}(z_j).
\end{equation}
From the unit circle, we now advance along the shape by expanding
around the next $z_j$, according to the scheme described above.

\begin{figure}
\centering
\epsfig{width=.40\textwidth,file=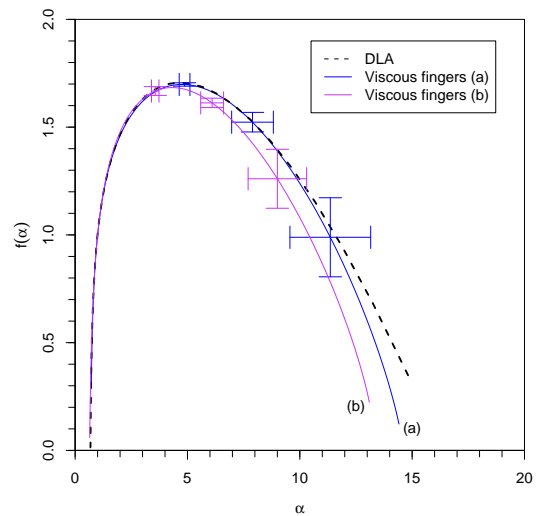}
\caption{$f(\alpha)$ curves for the harmonic measure of the experimental
  viscous fingers patterns and its comparison with the $f(\alpha)$
  curve of the harmonic measure of DLA. (a) is averaged over 15
  patterns obtained with a silicone oil with a
  dynamic viscosity $\mu=345$mPa$\cdot$s and (b) is averaged over 15
  patterns with $\mu=50.8$mPa$\cdot$s.}
  \label{spectrum}
  \end{figure}
\begin{figure}
\epsfig{width=.40\textwidth,file=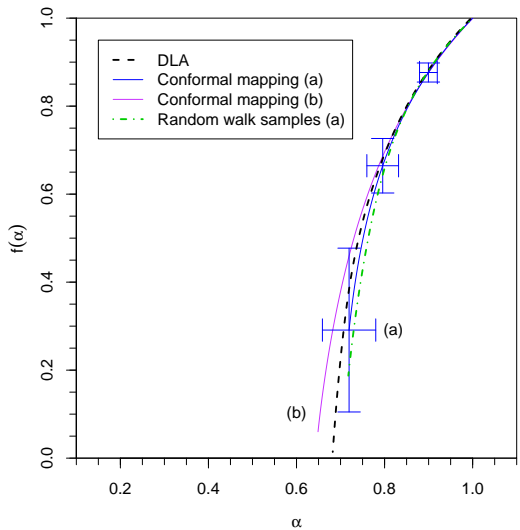}
\caption{Close-up of the $f(\alpha)$ curves presented in Fig.~\ref{spectrum} together with
the same curve computed from a sampling with random walkers. }
  \label{zoom}
\end{figure}

The efficiency of this method is surprising; the number of iterations
needed to build the conformal map remains rather low even for a
complex pattern like that in Fig.\ \ref{visfig}.  Figure \ref{map}
presents the result of our method using the digitalized image of
viscous fingers shown in Fig.\ \ref{visfig}.  Indeed, the conformal
map accurately reproduces the image. In general, the number of
iterations is around 10000-20000 for the maps that we analyze in this
Letter.

{\bf Results}: With the conformal mapping at hand, we have the
harmonic measure as $1/|\Phi^{'(n)} |$. One has to be cautious in
evaluating the derivative, since inside the fjords, the measure
becomes so small that it is difficult to resolve the fjords on the
unit circle. The way to evaluate the derivative involves a careful
book keeping of the positions of the individual bumps added in each
iteration step \cite{02JLMP}. In Fig.\ \ref{scaling} we present
log-log plots of the sum $\sum_{i=1}^{N(\epsilon)} P^q_i(\epsilon)$
vs. the box size $\epsilon$ for selected values of $q$. The estimated
slopes, $\tau(q)=(q-1)D_q$, are very reliable for positive values of
$q$. As seen in Fig.  \ref{scaling}, for values of $q<-0.25$, the
power-law scaling is lost.  In DLA, the deep fjords become dominant for
$q<-0.2$, and lead to a nonanalyticity in the $f(\alpha)$ curve, a
phenomenon known as a ``phase transition" in the thermodynamic
formalism of the fractal measure \cite{87KP}. 

We can now use the measured values of $\tau(q)$ to extract the
$f(\alpha)$ curve.  The result is shown in Fig. \ref{spectrum}
together with the same curve for the harmonic measure of DLA. While
there is no alternative way to compute this function in its entirety,
we can test the accuracy of our results by looking at $f(\alpha)$ for
$q>1$, which characterizes the highly probable regions of harmonic
measure near the tips.  We sampled the tips of the experimental data
with random walkers, a good method as long as one studies the part of
the multifractal spectrum for $q>1$. For the sampling, we use $10^6$
random walkers. Increasing this number did not lead to a significant
change in the results. Figure \ref{zoom} shows a close-up of the
$f(\alpha)$ curve presented in Fig. \ref{spectrum}, alongside the one
computed from the random walk samples. There is almost no difference
in the left hand branches, and we thus conclude that our method is
fully adequate for representing the tips of the viscous fingers.  The
estimate of $\alpha_{min}$ is the same (within the uncertainty of $\pm
0.06$) as the one computed for DLA, which was $\alpha_{min}=0.67$ \cite{03JMP}.

In conclusion, we have presented a powerful method for the
construction of a conformal map from the unit circle to complex
fractal patterns.  Using this conformal map, we can compute accurately
the fractal measure and determine its scaling properties.  The
comparisons shown in Figs. \ref{spectrum} and \ref{zoom} strongly
indicate that viscous fingering and DLA are in the same universality
class within the experimental uncertainty. Future study can determine
whether this correspondence extends to the nonanalyticity in the
$f(\alpha)$ curve.

\end{document}